\documentclass[aps,prc,showpacs,preprint]{revtex4-1}
\usepackage{graphicx}
\usepackage[section]{placeins}
\begin{document}
\title{THE QUANTUM HAMILTON JACOBI EQUATION AND THE LINK BETWEEN CLASSICAL AND QUANTUM MECHANICS}
\author{M. Fusco Girard}
\affiliation{Department of Physics ``E.R. Caianiello",\\University of Salerno \\and \\Gruppo Collegato INFN di Salerno,\\Via Giovanni Paolo II, 84084 Fisciano (SA), Italy\\
{\rm electronic address:} mario.fuscogirard@sa.infn.it}

\begin{abstract}
We study how the classical Hamilton's principal and characteristic functions are generated from the solutions of the quantum Hamilton-Jacobi equation. While in the classically forbidden regions these quantum quantities directly tend to the classical ones, this is not the case in the allowed regions. There, the limit is reached only if the quantum fluctuations are eliminated by means of coarse-graining averages. Analogously, the classical Hamilton-Jacobi scheme bringing to the motion's equations arises from a similar formal quantum procedure.
\end{abstract}

\maketitle

\section{Introduction}
Quantum Theory must approach Classical Theory asymptotically in the limit of large quantum numbers. This is equivalent to say that when $\hbar\to 0$, the laws of Quantum Mechanics (QM) must reduce to those of Classical Mechanics (CM). These are modern formulations of the Bohr's Correspondence Principle [1], assumed as a postulate of the Old Quantum Theory, and later confirmed in various aspects by the Quantum Mechanics. Intuitively, the principle is justified by an image like Fig. 1 and reproduced in many texts [2], where the quantum probability distribution function $|\psi(x)|^2$ for an high-level state of an harmonic oscillator is reported together with the probability distribution for the corresponding classical particle at the same energy. This latter quantity is defined as proportional to $1/v(x)$, where $v(x)$ is the particle's velocity. The figure suggests an empiric rule to obtain the classical quantity as the limit of the corresponding quantum one: firstly, consider large quantum numbers, and subsequently, eliminate the fluctuations by substituting some kind of averages to the exact values of the quantum function.

\begin{figure}[!h]
\includegraphics[scale=1]{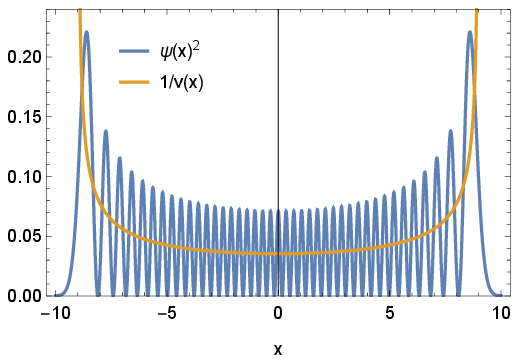}%
\caption{The comparison between the quantum probability distribution $\psi(x)^2$ (blue line) of the $n = 40$ state of a harmonic oscillator and the corresponding classical distribution (yellow) at the same energies. }
\end{figure} 

This empiric criterion is purely qualitative; moreover, the probability distribution is a basic concept in QM, but not in CM.  Finally, this classical probability diverges at the turning points. Therefore, we would like to have a more precise way to compare quantities, that are fundamental in both cases, and such that the classical quantity is the limit of the corresponding quantum one, for $\hbar \to 0$.
The Ehrenfest's theorem [3] states that the quantum expectation values of the coordinate and momentum operators evolve with time according to the classical Hamilton equation, if the force is replaced by its average. This seems a bridge between QM and CM. However, the theorem does not concern the limit $\hbar \to 0$, is true only for free particles or linear forces and, in general, it approximately holds only if the quantum fluctuation are small [4].
Between the various formulations of CM, the nearest one to the Schr\"odinger version of QM is based on the Classical Hamilton-Jacobi Equation (CHJE) [5]. The link is given by the Quantum Hamilton Jacobi Equation (QHJE), which appears looking for solutions of the Schr\"odinger equation in exponential form. The QHJE is the starting point for the WKB approximation [6]. Modern reviews of the WKB method (also named as JWKB or phase-integral method) are presented in [7, 8, 9 ].  In the framework of the usual Copenhagen interpretation of QM, the QHJE is fully equivalent to the Schr\"odinger equation, and reduces to the CHJE for $\hbar \to 0$.

As for this latter equation, its solutions are the Hamilton's principal and characteristic functions. These are fundamental quantities in CM, in that they allow or to completely solve the dynamical problem, in case of complete integrals [5], or to investigate the properties of families of trajectories, corresponding to special solutions [10-14].

Therefore, it seems natural to compare the solutions of the classical Hamilton-Jacobi equation, with the corresponding ones for the quantum case. This can be done for each number of degrees of freedom. For simplicity, we will consider here the one-dimensional conservative case. The QHJE appears when the particle's wave function at the energy E in a potential $V(x)$ is searched in the form:
\begin{equation}
\psi(x,E,t)=A\ e^{{i\over \hbar}S{x,E,t}}
\end{equation}    
where $S(x, E, t)$ is a complex quantity, and A is a constant.
When Eq. (1) is inserted into the Schr\"odinger equation:
\begin{equation}
i\hbar{\partial\psi\over\partial t} = \left[-{\hbar^2\over 2m}{\partial^2\psi\over \partial x^2}+V(x)\right]\psi\ ,
\end{equation}     
the QHJE results:
\begin{equation}
-{\partial S \over \partial t}= {1\over 2m}\left[\left( {\partial S\over \partial x}\right)^2 -i\hbar {\partial^2 S\over \partial x^2}\right] - V(x)\ .
\end{equation}				

The time dependence can be separated, by writing:
\begin{equation}
S(x,E,t) = W(x,E) - E t\ ,
\end{equation}  
Then, Eq. (3) becomes the time-independent QHJE:
\begin{equation}
{1\over 2m}\left({dW\over dx}\right)^2 - {i\hbar \over 2m}{d^2W\over dx^2}=E - V(x)\ .
\end{equation}  

Like $S(x, E, t)$, the function $W(x, E)$ is in general a complex quantity.

In (5), the energy $E$ is considered as a fixed parameter, and therefore according to the common usage, the derivatives with respect to $x$ are written as ordinary derivatives. In this paper, these will be usually indicated by means of primes (i.e.  ${dW \over dx}=W'(x,E)$). In the last section of this paper, we will need also the derivatives of $W(x, E)$ and $S(x, E, t)$ with respect to $E$.

By setting $\hbar = 0$, the equations (3) and (5) become the classical time-dependent and time-independent Hamilton-Jacobi equations, respectively, whose solutions $S_C(x,E,t)$ and $W_C(x,E)$ are the Hamilton's principal and characteristic functions [5], also named the action and abbreviated action, respectively [15].

Therefore, a solution $S(x, E, t)$ of the QHJE (3) will be in analogy called quantum Hamilton's principal function (or quantum action), and a solution $W(x, E)$ of the Eq. (5) will be called quantum Hamilton's characteristic function, (or quantum abbreviated action).

In order to investigate how the classical mechanics arises in this approach from the quantum one, is seems natural to compare these quantum actions with the corresponding classical quantities.

This problem was already touched in [16], according to a method that for some aspects can be considered as the exact version of the WKB approach. The aim of the present paper is to more completely investigate this point.

The usual WKB method constructs approximate solutions of Eq. (5), by expanding $W(x, E)$ in powers of $\hbar$, and neglecting terms of higher orders than $\hbar ^2$. The resulting semi classical wave function has an exponential expression in the classically forbidden region (c.f.r.), and a trigonometrical one in the allowed region (c.a.r.). It usually fits very well the exact wave function, except near the turning points, where diverges.

The method presented in [16] differs from the WKB one in that it makes use of exact solutions of Eq. (5). In this way, the wave functions are precisely represented along the entire $x$-axis, turning points included. In the following, we briefly resume the method, referring to the quoted references for the details.

\section{THE METHOD}
The Eq. (5) is a second order non-linear equation for $W(x, E)$, but it can also be seen as a first order equation for the derivative $W'(x, E)$. This quantity was named by Leacock and Padgett [17, 18] the quantum momentum function $p(x, E)$. It is an ordinary complex function, not to be confused with the quantum operator momentum, which does not appear in the following:
\begin{equation}
 p(x,E) = W'(x,E) = {\hbar\over i} {\psi '(x,E)\over \psi(x,E)}\ .
\end{equation} 
With this definition, (5) becomes a Riccati equation for $p(x, E)$:
\begin{equation}
 i\hbar p' = (p)^2- 2m (E-V(x))\ .
\end{equation}    
Leacock and Padgett demonstrated that the exact quantum energy levels can be obtained, without solving (7), from the condition:
\begin{equation}
\oint p (x,E) dx = 2n\pi\hbar\ ,
\end{equation}                
where the integration is done along a closed path in the complex x-plane, enclosing the turning points.

Found $p(x, E)$ from (7), the solution of (5) is:
\begin{equation}
W(x,E) = \int p(x,E) dx\ .
\end{equation} 

By setting  $\hbar= 0$, the Eq. (7) becomes the equation for the classical momentum $p_C$:
\begin{equation}
p_c(x,E) = \pm \sqrt{2m(E-V(x))}\ ,
\end{equation}    
whose integration gives the classical abbreviated action:
\begin{equation}
W_C(x,E)=\int p_C(x,E)dx
\end{equation}               
When $\hbar = 0$, the QHJE becomes the CHJE, and therefore the quantum abbreviated action $W (x, E)$ generates in some way the classical corresponding one $W_C(x, E)$. Similarly, its derivative, i.e. the quantum momentum function $p(x, E)$, has to become the classical momentum $p_C(x, E)$.

The quantum abbreviated action $W(x, E)$ is therefore the suitable quantity to investigate, being the fundamental function in the QHJ formulation of QM. Indeed, from it, the quantum action $S(x, E, t)$ is obtained by means of Eq. (4), and subsequently, the wave function is given by Eq. (1).

As discussed in [16], the Eqs. (5) and (7) admit many solutions, generating the same wave function through Eqs. (4) and (1). The simplest, special ones, are obtained as shown in [19], by analysing the polar structure of (7). We will indicate these special solutions as $W_S (x, E)$ and $p_S (x, E)$, respectively. For low-lying states, these solutions can often be found by simple inspection. For instance, it is immediate to verify that a special solution of (7) for the ground state of a harmonic oscillator of mass $m$ and frequency $\omega$ is $p_S = i m\omega x$, with the corresponding action $W_S  = {1\over 2}  i m\omega x^2$.

The special solutions so found are the same as obtained from the complex logarithm of the wave function, by means of Eqs. (1) and (4).

As shown in [16], in the forbidden regions the imaginary classical actions are the limits of the special solutions $W_S(x, E)$ and $p_S(x, E)$ for $\hbar \to 0$. In the classically allowed regions instead, when we try to connect these solutions with the corresponding classical quantities, we immediately run into serious difficulties.

In the absence of magnetic field, the wave functions can be taken as real [20]. Then, the quantum momentum function $p_S (x, E)$, as computed from (6) is a purely imaginary quantity (more exactly, it is a complex quantity with a real part everywhere zero, apart from delta singularities at the nodal points of the wave function, see below). The classical momentum instead is imaginary inside the classically forbidden regions, but it is real in the classically allowed ones. Analogously, the reduced quantum action $W_S(x, E)$, as computed from (1), is a complex quantity, with an imaginary part logarithmically diverging at the nodal points, and a real part discontinuously jumping from 0 to $\pi\hbar ({\rm mod} \pi\hbar)$ at every variation in sign of the wave function (these jumps produce the delta singularities of ${\rm Re} [p_S(x, E)]$). The classical reduced action $W_C(x, E)$ is instead a continuous real function inside the classically allowed regions, and imaginary outside.

Therefore, $W_C (x, E)$ and $p_C (x, E)$ inside the classical regions, cannot be the limits of the special solutions $W_S (x, E)$ and $p_S (x, E)$, but have instead to be generated by the real parts of two complex continuous functions whose imaginary parts vanish when $\hbar \to 0$. As shown in [16], these functions can be obtained from the general solutions $W_G(x, E)$ and $p_G(x, E)$ of the equations (5) and (7).

According to the previous considerations, as in the usual WKB method, we have to differently treat Eq. (5) in the classically forbidden and allowed regions. Let us consider the simplest case of a potential $V(x)$ such that the classical region, indicated as II, is located between the turning points $x_1$ and $x_2$. The forbidden regions $x < x_1 $  and  $x > x_2$ are indicated as I and III, respectively. The energy eigenvalues come from the condition (8) and the corresponding  imaginary quantum momentum functions $p_{S,I,III} (x, E)$ in I and III are found from (7) as explained above. By integrating, we get:
\begin{equation}
W_{S,I,III}(x,E) = i Y_{S,I,III} (x,E) = \int p_{S,I,III} (x, E) dx\ .
\end{equation}   
Therefore, in the forbidden regions the time-independent wave functions have the respective exact exponential WKB-like representations:
\begin{equation}
\psi_{I,III}(x,E)=A_{I,III}e^{-Y_S(x,E)/\hbar}= A_{I,III}e^{-{1\over\hbar }\int p_S (x,E)dx}\ .
\end{equation}   
$A_{I,III}$ are constants, to be fixed later.
In the classical region, we need instead the general complex solution of Eq. (5), of the form:
\begin{equation}
W_G(x) =X(x) + i Y(x)\ .
\end{equation}             

This general solution $W_G(x, E)$ can be built starting from the special solutions $W_S(x, E)$ and $p_S(x, E)$, by applying a known theorem for the Riccati equation [21]. It states that if one special solution $p_S (x, E)$ of (7) is known, the equation can be completely integrated and the general solution is given by:
\begin{equation}
p_G (x) = p_S (x) + {1\over v(x)} \ ,
\end{equation} 
where $v(x, E)$ is the general solution of an associated linear differential equation, which in our case is:
\begin{equation}
v'(x) - \left({2i\over \hbar}\right) p_S (x) v(x) ={i\over \hbar}\ .
\end{equation}       

The result is:
\begin{equation}
p_G(x) = p_{S}(x) + {e^{-{2i\over\hbar} \int_0^x p_{S}(x) dx} \over {i\over \hbar}\int _0^x e ^{-{2 i\over \hbar } \int_0^x p_{S}(x) dx} dx +C_0}\ ,
\end{equation} 
whose integration gives:
\begin{equation}
W_{G}(x) = W_{S}(x) +{\hbar\over i} \log \left[{i\over \hbar}\int_0^x  e^{-{2i\over\hbar} \int_0^x p_{S}(x) dx} dx + C_0\right] + C_1\ .
\end{equation}   
$C_0$ and $C_1$ are two complex constants.
The real part of (18) is:
\begin{equation}
X(x) = Re[W_S(x)] + \hbar Arg \left[{i\over \hbar}\int_0^x  e^{-{2i\over\hbar} \int_0^x p_{S}(x) dx} dx + C_0\right] + Re [C_1]\ .
\end{equation}   
For various potentials, the integrals in (17) and (18) can be analytically done [16]. For instance, the special solution of (7) for the quantum momentum function of the n state of the harmonic oscillator (ho) is [19]:
\begin{equation}
p_{F,S}^{ho,n} (x) = i \left( m\omega x -{2n\sqrt{m\omega\hbar}H_{n-1}\left(\sqrt{m\omega\over\hbar}x\right) \over H_{n}\left( \sqrt{m\omega\over\hbar}x\right)	} \right)\ ,
\end{equation}    
where $H_n$ is the $n$-th Hermite polynomial.

Therefore, according to (17), the corresponding general solution is:
\begin{equation}
p_{G}^{ho,n} (x) = p_{S}^{ho,n} (x)  + {e ^{m\omega x^2\over \hbar} \over H_{n}^2\left(\sqrt{m\omega\over\hbar}x\right) \left[{i\over \hbar }\int_0^x{{e ^{m\omega x^2\over \hbar} \over  H_{n}^2\left(\sqrt{m\omega\over\hbar}x\right)}}dx + C_0^{ho,n}\right]} \ .
\end{equation}      
By integrating (20) and (21) one obtains, respectively, the special solution of (5):
\begin{equation}
W_{S}^{ho,n} (x) = i Y_{S}^{ho,n}(x) = i \left( {1\over 2} m\omega x^2 -\hbar \log \left[ H_{n}\left(\sqrt{m\omega\over\hbar}x\right)\right]	\right)\ ,
\end{equation}   
apart for an unessential integration constant, and the corresponding general one:
\begin{equation}
W_{G}^{ho,n} (x) = W_{S}^{ho,n}(x) +{\hbar\over i} \log \left[ {i\over \hbar} \int_0^x{{e ^{m\omega x^2\over \hbar} \over  H_{n}^2\left(\sqrt{m\omega\over\hbar}x\right)}}dx + C_0^{ho,n}\right]+ C_1^{ho,n}  \ .
\end{equation}   
The C's are constants.
The real part of last expression is:
\begin{equation}
X^{ho,n} (x) = \hbar\ Arg \left[H_{n}\left(\sqrt{m\omega\over\hbar}x\right)\right] +\hbar\ Arg \left[ {i\over \hbar} \int_0^x{{e ^{m\omega x^2\over \hbar} \over  H_{n}^2\left(\sqrt{m\omega\over\hbar}x\right)}}dx + C_0^{ho,n}\right]+ Re[C_1^{ho,n}] \ .
\end{equation}   
The real part of $p_G(x, E)$ can analogously be computed from (21).

For each value of the integer $n$, the integrals in (21)-(24) can be analytically done, and the results can be expressed in terms of elementary function and the error function of imaginary argument, which is connected to the Dawson integral [22].

For a general Hamiltonian, by inserting in (1) the $W_G (x, E)$ analytically or numerically computed from Eq. (18), with the constants $C_0$ and $C_1$ chosen as described in [16],  one has the exact solution of the Schr\"odinger equation in the classical region.

The same results can be obtained by means of a different, mainly numerical procedure [23].

In the classically allowed region, when (14) is put into (5) and the real and imaginary parts are separated, the following equations are obtained for the real X(x, E) and the imaginary part $Y(x, E)$ of $W(x, E)$ (the dependence on $E$ here and in the following equations will be understood):
\begin{equation}
X'^2(x) - Y'^2(x) + \hbar Y''(x)  = 2m\left(E -V(x) \right)
\end{equation}   
\begin{equation}
X'(x) Y'(x) -{1\over 2} \hbar X''(x)  = 0 \ .
\end{equation}   
Last equation gives:
\begin{equation}
Y(x)=\hbar \log\left[\sqrt | X'(x)|\right] \ .
\end{equation}   
By putting (27) into (25), the following equation results:
\begin{equation}
 {4X'^4(x)-3\hbar ^2 X''^2(x) + 2\hbar^2X'(x)X'''(x)\over 4X'^2(x)} =2m(E-V(x))\ .
\end{equation}       
This third order differential equation is rigorously equivalent to the Schr\"odinger equation [4, pag. 232].

When $\hbar = 0$, last equation becomes the CHJE for $X(x, E)$, while $Y(x, E)$ vanishes according to (27). This confirms that in the c.a.r. the classical reduced action $W_C(x, E)$ is generated, in the classical limit, by the real part $X(x, E)$ of the quantum action $W(x, E)$, as claimed above.

With the suitable Cauchy data [23], the non-linear equation (28) can be numerically integrated, giving the same results as (19). The solution is a continuous function $X(x, E$), different from the step function which is the real part of the special solution $W_S (x, E)$.  By putting it and $Y(x, E)$ from Eq. (27) into (14), one obtains the quantum action $W(x, E)$, and from this latter, the time independent complex wave function (1):
\begin{equation}
{A\over \sqrt{X'(x)}}\exp\left[{i\over\hbar}X(x)\right]
\end{equation}                
with a complex constant A.
The Eq. (28) is equivalent to the Eq. (3.6) of Ref. [7], which is written in a different form and obtained through another approach, and (3.7) there is equal to (29). There too it is claimed that knowing any solution of (28), one has the exact solution (29) of the Schr\"odinger equation, but no attempt is done to get this solution.

By suitably choosing the constants and combining (29) and its conjugate, the wave function in the classically allowed region can be written in the WKB like form:
\begin{equation}
\psi_{II}(x)=\frac{A_{II}}{\sqrt{|X'(x)|}}\sin\left[{X(x)\over \hbar}+ {\pi \over 4}\right]
\end{equation} 		     
where $A_{II}$ is a real constant.

The constant $\pi\over 4$ in (30) is chosen in order to put the wave function in the WKB-like expression. This latter has the classical reduced action $W_c(x, E)$ in place of the quantum function $X(x, E)$.

The comparison between (30) and the corresponding WKB expression confirms that the real part X(x, E) of the quantum reduced action generates the classical reduced action $W_C(x, E)$ in the limit $\hbar \to 0$, and its derivative $X'(x, E)$ generates the classical momentum $p_C(x, E)$.

In the classically forbidden regions, outside the turning points, the wave function has instead the exponential representation (13), but this time the functions $Y_{I,III} (x, E)$ are numerically computed from the Eq. (25), with $X(x) = 0$:
\begin{equation}
- Y'^2(x) + \hbar Y''(x)  = 2m\left(E -V(x) \right)
\end{equation}   
When $\hbar = 0$, this equation reduces to the CHJE in the forbidden region.

The real constants $A_i$ in (13) and (30) are to be fixed by the continuity of the wave function and its first order derivative at the turning points. The numerical version of the method is independent from the analytic one, and allows finding the allowed energy values too.  Indeed, in [24] it has been shown that a value of the parameter E in (28) and (31) is an energy eigenvalue, if with this choice it is possible to construct a normalizable wave function, continuous with its derivative, by matching together at the turning points the functions (13) and (30), by a suitable choice of the $A_i$. Obviously, this is the usual quantization condition for the Schr\"odinger equation.

This method to find the energy eigenvalues has been successfully applied to various Hamiltonians, and gives the eigenvalues with the same precision as the usual approaches [24, 25].

The Eqs. (13) and (30) give a WKB-like representation of the wave function along the entire x-axis. However, (13) and (30) are exact, and exactly reproduce the wave function at the turning points too, where the WKB expressions diverge. Moreover, it is important to note that the representation (30) of the wave function in the c.a.r. is not possible by using the real part of the special solution $W_S(x, E)$, which is a step function.

The Eq. (30) shows that the real part $X(x, E)$ of the quantum reduced action is a fundamental quantity in QM, being the phase of the wave function in the classical region, while its derivative $X'(x, E)$ controls the amplitude.

The detailed study of the solutions of our equations for the harmonic oscillator and the hydrogen atom is given in [16, 23] and for the quartic oscillator in [24], and will not be repeated here. We simply present in Fig. 2 a graph comparing the real part $X(x, E)$ of the quantum reduced action for the state $n = 2$ of the harmonic oscillator, with the corresponding classical quantity $W_C(x, E)$, at the same energy.

\begin{figure}[!h]
\includegraphics[scale=1]{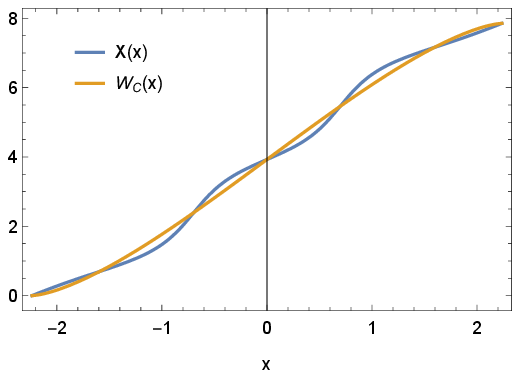}%
\caption{The real part $X (x, E)$ of the quantum abbreviated action for the state $n = 2$ of a harmonic oscillator (blue) and the corresponding classical quantity $W_C(x, E)$
 (orange). The figure refers to a semi-period of oscillation of the classical particle, from the left turning point to the right one. }
\end{figure} 

The two functions refer to a semi-period of oscillation of the classical particle, from the left turning point $x_1$ to the right one, $x_2$. As the equation (28) does not contain $X(x, E)$, but only its derivatives, a constant can be added to $X(x, E)$, and the same holds for $W_C(x, E)$. Therefore, the value of the two functions in $x_1$ is arbitrary and is chosen equal to 0. The choice of the other two conditions needed to solve the Cauchy problem for the Eq. (28) is explained in the quoted references. As seen from the figure, both the functions $X(x, E)$ and $W_C(x, E)$, are monotonically increasing from the value 0 in $x_1$ to $(n +1/2) \pi\hbar$ in $x_2$. The quantum function follows the profile of the classical one, waving around it, and the number of ripples increases with $n$. These ripples cause peaks in the real part $\Re (p(x, E)) = X'(x, E)$, as seen from Fig. 3.

\begin{figure}[!h]
\includegraphics[scale=1]{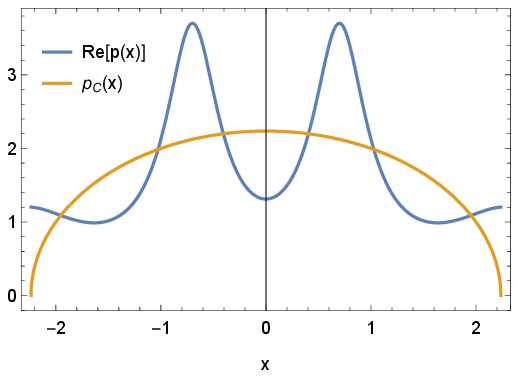}%
\caption{The real part $p (x, E)$ of the quantum momentum function for the state $n = 2$ of a harmonic oscillator (blue) and the corresponding classical momentum $p_C(x, E)$ (orange). The figure refers to a semi-period of oscillation of the classical particle, from the left turning point to the right one. }
\end{figure} 

In Fig. 4 are plotted the functions $\sin [X^{ho,n}(x)/\hbar+ \pi/4]$ (green line), $1/\sqrt{X'(x)}$ (orange line) and finally their product (blue line) which according to eq. (30), exactly reproduces the $n= 2$ wave function for the harmonic oscillator.

\begin{figure}[!h]
\includegraphics[scale=1]{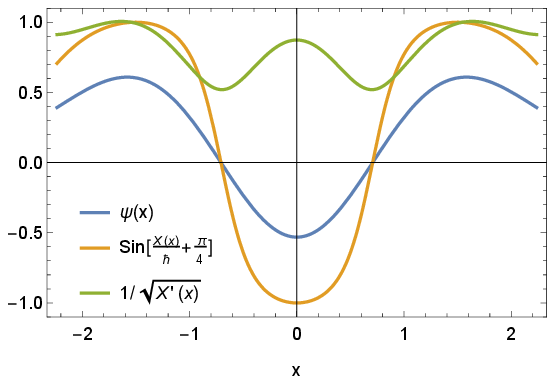}%
\caption{The yellow line is the function $\sin[X^{ho}(x)/h + \pi/4]$. The green line is the function $1/\sqrt{X' (x)}$. Their product gives the wave function for the $n = 2$ state of the harmonic oscillator, inside the classically allowed region (blue line). }
\end{figure} 

\section{THE CLASSICAL LIMIT}

In order to investigate the classical limit, it is again necessary to distinguish the classically forbidden regions from the allowed one.

As for the c.f.r., the special solutions $W_S(x, E)$ and $p_S(x, E)$ directly generate the classical reduced action and momentum, respectively. For a generic hamiltonian, this can be seen from the Eq. (31) which reduces to the CHJE (with $V(x) > E$), when $\hbar = 0$. The way in which the limit is approached for the harmonic oscillator can be seen from (22), to be compared with the classical action $W_C(x, E)$:
\begin{equation}
W_C^{ho,n}(x,E) = i\left({1\over 2}x\sqrt{m^2\omega^2x^2-2mE} - {E\over\omega}\log\left[m\omega^2x+\omega\sqrt{m^2\omega^2x^2-2mE}\right]\right)\ .
\end{equation}					
To both the actions (22) and (32), a constant can be added, so that the two functions can be chosen as equal for a particular value of $x$.  The numerical values of the imaginary parts of the two functions are plotted in Fig. 5, which refers to $n = 20$.
\begin{figure}[!h]
\includegraphics[scale=1]{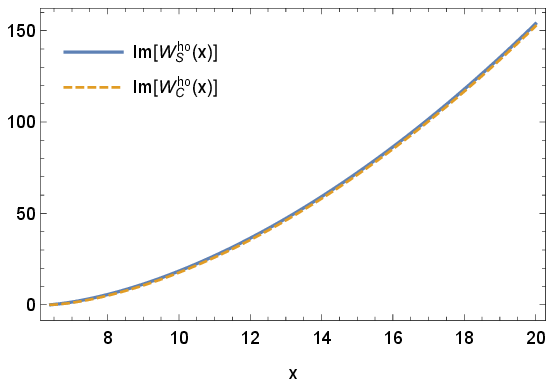}%
\caption{The imaginary part of the quantum abbreviated action $W^{ho}S (x,E)$ in the classically forbidden region III (blue line), for the $n = 20$ state of the harmonic oscillator, as compared with the corresponding classical quantity $W^{ho}_C (x,E)$ (dashed, orange). The two curves are practically superimposed. }
\end{figure} 

As seen from the figure, the numerical values are very close, and their relative difference tends to vanish for large $x$. This implies that the classical imaginary momentum too in the c.f.r. is generated by the quantum momentum function as given by the Eq. (20).

 As for the c.a.r., we note that in the Eqs. (16) and (17), the dependence on $\hbar$ is non-analytical, therefore, the expansion in power series of this quantity is not possible. Anyway, a clear indication of what happens in the limit can be obtained by using numerical computations with increasing values of the quantum number $n$. Some results are presented in Fig. 6, where the real part of the quantum abbreviated function for the harmonic oscillator with $n = 60$, is reported.  As seen from the figure, while increasing $n$, the real part $X(x)$ of this function seems to tend more and more in this scale to the classical action $W_C(x)$. Actually, however, it maintains a waving behavior around this latter, so acquiring in the limit an infinite number of ripples. The oscillations' amplitude tends to become constant while increasing n, while their number increases. This can be seen from Fig. 7, where the difference $X(x, E) - W_C(x,E)$ is plotted.
\begin{figure}[!h]
\includegraphics[scale=1]{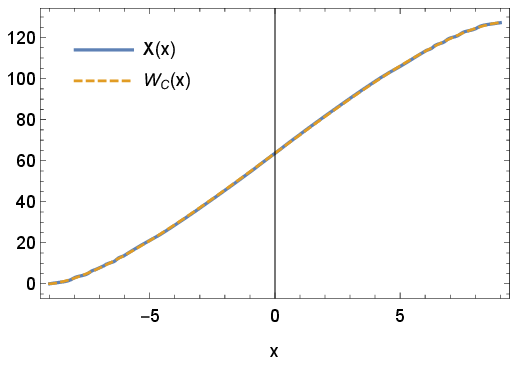}%
\caption{The real part $X (x, E)$ of the quantum abbreviated action for the state $n = 60$ of a harmonic oscillator (blue) and the corresponding classical quantity $W_C(x, E)$ (dashed, orange). The figure refers to a semi-period of oscillation of the classical particle, from the left turning point to the right one. The two functions are chosen equal to 0 in the left turning point. In the scale of the figure, the two curves seem overlapping, but in reality the quantum one waves around the classical. }
\end{figure} 

\begin{figure}[!h]
\includegraphics[scale=1]{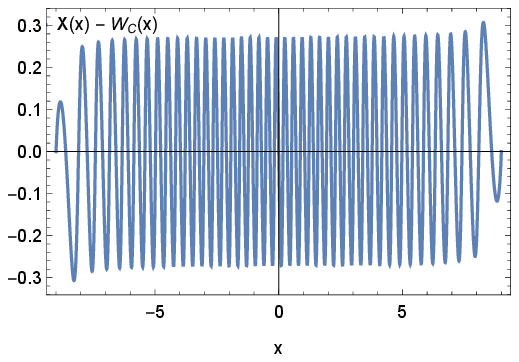}%
\caption{The difference $X (x, E) - W_C(x, E)$ between the real part $X (x, E)$ of the quantum abbreviated action for the state $n = 60$ of a harmonic oscillator and the corresponding classical quantity $W_C(x, E)$, in the classically allowed region. While increasing n, the amplitude of the oscillations tends to become constant, and their number increases.}
\end{figure} 
As for the quantum momentum function, its real part ${\rm Re}[p_G(x)$, which is reported in Fig. 8 for $n = 60$, presents oscillations of finite heights, due to the ripples in the real part of the quantum abbreviated action.
 \begin{figure}[!h]
\includegraphics[scale=1]{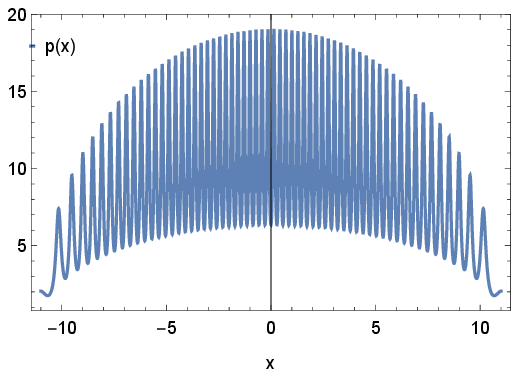}%
\caption{The real part $p (x, E)$ of the quantum momentum function for the state $n = 60$ of a harmonic oscillator, in the classically allowed region. }
\end{figure} 
The number of these oscillations increases with n, as can be seen by comparing this figure with the Fig. 3, which refers to $n = 2$. The presence of a number of peaks and oscillations going to infinite, demonstrates that the quantum functions cannot directly tend in strict mathematical sense to the corresponding classical quantities. The figures however suggest investigating what happens if the oscillations are eliminated by means of a coarse graining procedure. This means to divide the interval between the turning points in a number of sub-intervals, and in each of these the average value is substituted to the exact values of the functions.  
\begin{figure}[!h]
\includegraphics[scale=1]{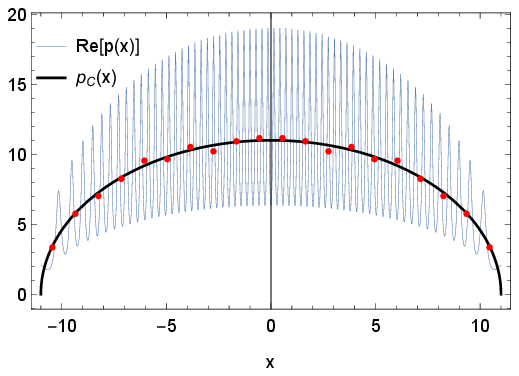}%
\caption{The red points represent the coarse-grained averages of the real part $p (x, E)$ of the quantum momentum function for the state $n = 60$ of a harmonic oscillator, in the classically allowed region. The black line is the corresponding classical momentum $p_C(x, E)$.}
\end{figure} 
In Fig. 9, the result of such operation on the real part of $p_G(x)$ for $n = 60$ is plotted: the red dots represents the mean values of this function, computed averaging it in 20 subintervals of the x-axis, between the turning points. As seen from the figure, the dots are distributed along the curve of the classical momentum $p_C(x)$, which is represented by the black line. As for the imaginary part of the quantum momentum function, it symmetrically oscillates around zero, so that its mean values in each small subinterval tends to zero; in addition, it is also proportional to $\hbar$ according to Eq. (27), and therefore vanishes in the classical limit. The quantum momentum function for $\hbar \to 0$ in the c.a.r. so becomes purely real, and generates the classical momentum if its exact values are averaged by means of the coarse-graining. The same happens to its integral, i.e. the quantum abbreviated action. Similar computations for various hamiltonians show the same behavior. The diverging number of fluctuations in the limit $\hbar \to 0$  explains why the WKB series expansion does not converge to the quantum characteristic function.

An analogous investigation can be done in order to see if the classical equations of motion are generated by a sort of quantum counterpart.

In the Hamilton-Jacobi formulation of the Classical Mechanics, as well known, the relation between $x$ and $t$, i.e. the motion's equation in the form $t = t (x)$, is obtained by equating to a constant $\beta$ the derivative of the action $S_C(x, E, t)$ with respect to the energy $E$ [5]. This procedure is usually considered as the result of a canonical transformation to a null Hamiltonian function. However, the same equation appears by separating the variables and integrating the equation expressing the energy conservation.
For a semi period of the harmonic oscillator, this gives:
\begin{equation}
{1\over \omega}\arcsin \left[{\sqrt m\omega^2} \over \sqrt{2E} x \right] = t+\beta
\end{equation} 			
By inverting this equation one obtains the usual form of the motion's equation $x = x (t, E, \beta)$.
In order to see what happens by formally applying this classical procedure to QM, we have to derive the quantum action $S(x, E, t)$ with respect to the energy E:
\begin{equation}
{\partial S(x,E,t) \over \partial E} = {\partial W (x,E)\over \partial E}-t\ .
\end{equation}                   
It is possible to obtain a linear differential equation for this quantity, by deriving the Riccati equation (7) with respect to E. The result however contains integrals of the quantum momentum function (22), which is already given by a complicated expression. The final formulae are therefore too cumbersome to be useful, so that we prefer to adopt the numerical procedure.

We already know that in the classical limit, the complex quantum quantities in the c.a.r. become real functions. Therefore, we only need the derivative with respect to the energy E of real part $X(x, E)$ of the quantum abbreviated action $W(x, E)$. We will indicate this derivative with the subscript E, i.e:
\begin{equation}
X_E(x,E)={\partial  X(x,E)\over \partial E}\ .
\end{equation}            
The equation for this quantity is obtained by deriving Eq. (28) with respect to E. The result is:
\begin{eqnarray}
4X'(x)^4 X'_E(x) + && 3\hbar^2 X'_E(x) X'' (x)^2 - \hbar^2 X'(x)  \nonumber\\
&& \left( 3X''(x) X''_E(x) + X'_E(x) X'''(x)\right)+\hbar^2 X(x)^2X'''E(x) = 4mX'(x)^3\ .
\end{eqnarray}	
In this equation, the explicit dependence on E is understood, and the primes indicate the derivatives with respect to x, as elsewhere in this paper. The Eq. (36) is a differential equation for $X_E(x, E)$, giving the dependence of this quantity on the coordinate $x$, at fixed energy. The function $X(x, E)$ here and its derivatives with respect to x, are computed by previously integrating the Eq. (28).

The Fig. 10 reports the typical results of the numerical integration of (36). The figure refers to the state with quantum number n = 50 of the harmonic oscillator. The curve is the derivative $X_E (x, E)$ of the real part $X(x, E)$ of the quantum abbreviated action $W (x, E)$, with respect to the energy E. As seen from the figure, this derivative is a highly oscillating function, whose oscillations are contained between two monotonic increasing functions. The number of these oscillations goes to infinity when increasing n.
\begin{figure}[!h]
\includegraphics[scale=1]{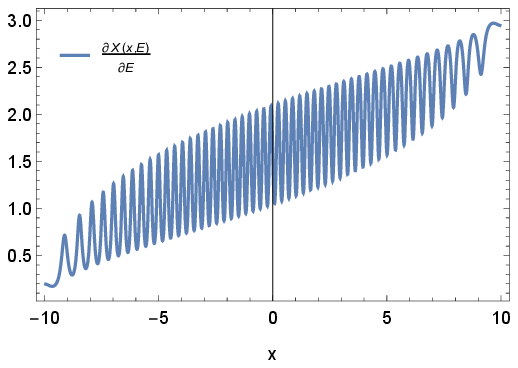}%
\caption{The derivative with respect the energy E of the real part of the quantum abbreviated action  for the $n = 50$ state of the harmonic oscillator, as computed by the numerical integration of the Eq. (36). }
\end{figure} 
From the figure it is clear that equating to a constant the r.h.s. of (34) gives a multi-valued relation between $x$ and $t$.  Indeed, graphically it means to find the intersections between the graph in the figure and the horizontal lines with $y$-coordinates $t+\beta$.
\begin{figure}[!h]
\includegraphics[scale=1]{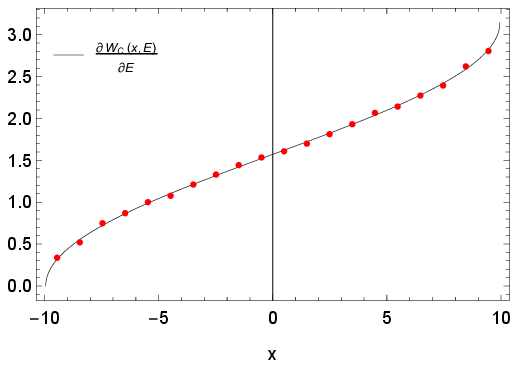}%
\caption{The red points are the coarse-grained averages of the function reported in Fig. 10. The black line is the derivative with respect to the energy of the classical abbreviated function $W_C(x, E)$.}
\end{figure} 
 However, if we eliminate the oscillations by means of a coarse-graining, we obtain the red points in Fig. 11. The black curve represents the derivative $\partial W_C (x,E)\over \partial E$ of the classical abbreviated action $W_C(x, E)$ for the harmonic oscillator, given by the l.h.s. of (33). As seen from the figures, the averaged quantum values follow quite well the corresponding classical curve. Therefore, the formal application of the classical procedure to the quantum action, going to the classical limit after the elimination of the quantum oscillations by means of the coarse graining, gives the classical equation of motion.

The results presented show that the fundamental quantities of the Hamilton-Jacobi formulation of Classical Mechanics, i.e. the action and the abbreviated action, are generated in the c.a.r. by the real parts of the corresponding quantum functions, while the imaginary parts vanish. The quantum functions in these regions present a number of oscillation increasing with the quantum number n, and going to infinity in the classical limit. Therefore, the classical quantities in the c.a.r. cannot be the limit of the quantum corresponding quantities, in the strict mathematical sense. In order to obtain the classical quantities from the quantum ones, these latter have to be previously smoothed by eliminating the quantum oscillations by means of the coarse graining averages, and are these smoothed functions that tend to the classical ones. The same happens for the quantum momentum function, which generates the classical momentum. The classical equation of motion arises in an analogous way, from the formal multivalued quantum relation, obtained by applying to the quantum action the classical Hamilton-Jacobi procedure.

As the true description of the motion is given by the quantum mechanics, it is clear that the macroscopic objects apparently follow the laws of the classical mechanics due to the fact that the macroscopic measure instruments, perform a coarse-graining averages, eliminating the intrinsic quantum oscillations.
The empiric rule inferred by Fig. 1 is in this way confirmed and clarified.

\end{document}